# Robust Nucleus Detection With Partially Labeled Exemplars

LINQING FENG , (Member, IEEE), JUN HO SONG, JIWON KIM, SOOMIN JEONG,
JIN SUNG PARK, AND JINHYUN KIM
Center for Functional Connectomics, Brain Science Institute, Korea Institute of Science and Technology (KIST), Seoul 02792, South Korea

Corresponding authors: Linqing Feng (feng@kist.re.kr) and Jinhyun Kim (kimj@kist.re.kr)

This work was supported in part by the Korea Institute of Science and Technology (KIST) Institutional Program under Project 2E27850, and in part by the National Research Foundation of Korea (NRF) Brain Research Program under Grant NRF-2017M3C7A1043838.

**ABSTRACT** Quantitative analysis of cell nuclei in microscopic images is an essential yet challenging source of biological and pathological information. The major challenge is accurate detection and segmentation of densely packed nuclei in images acquired under a variety of conditions. Mask R-CNN-based methods have achieved state-of-the-art nucleus segmentation. However, the current pipeline requires fully annotated training images, which are time consuming to create and sometimes noisy. Importantly, nuclei often appear similar within the same image. This similarity could be utilized to segment nuclei with only partially labeled training examples. We propose a simple yet effective region-proposal module for the current Mask R-CNN pipeline to perform few-exemplar learning. To capture the similarities between unlabeled regions and labeled nuclei, we apply decomposed self-attention to learned features. On the self-attention map, we observe strong activation at the centers and edges of all nuclei, including unlabeled nuclei. On this basis, our region-proposal module propagates partial annotations to the whole image and proposes effective bounding boxes for the bounding box-regression and binary mask-generation modules. Our method effectively learns from unlabeled regions thereby improving detection performance. We test our method with various nuclear images. When trained with only 1/4 of the nuclei annotated, our approach retains a detection accuracy comparable to that from training with fully annotated data. Moreover, our method can serve as a bootstrapping step to create full annotations of datasets, iteratively generating and correcting annotations until a predetermined coverage and accuracy are reached. The source code is available at https://github.com/feng-lab/nuclei.

## I. INTRODUCTION

The cell nucleus is a fundamental biological structure containing important information, such as the cell type, density, and viability. Automatically identifying and segmenting nuclei from microscopic images is the first step in many types of quantitative analyses with applications ranging from basic cell biology [1], [2] and systems neuroscience [3], [4] to cancer diagnosis [5]. This task is a challenging instance-segmentation task because it requires the correct detection of all instances of nuclei within an image, along with precise labeling of the pixels belonging to each detected nucleus. Nucleus segmentation methods need to be instance-aware to correctly separate touching nuclei.

Meanwhile, such methods need to address difficulties such as a high cell density, low contrast, intensity inhomogeneity, shape variation, weak boundaries, strong gradients inside the nucleus, and ambiguous overlapping.

Classical approaches, including thresholding [6]–[10], marker-controlled watersheding [11]–[13], edge detection [14], shape matching [15]–[17] and region merging/growing [18]–[22], often assume a certain signal pattern of nuclei or cells, such as bright centers, strong boundaries or low signal gradients inside the nuclei. These methods work well for certain datasets but tend to fail in difficult cases in which the assumptions do not hold. Denoising and transformation [9], [13] could be used to improve these methods by creating transformed images that fit the assumptions more closely. Nevertheless, the applicability of such methods is limited by these assumptions.

  

On the other hand, machine learning-based approaches can learn nuclear signal patterns directly from labeled examples without any prior assumptions. Indeed, with sufficient training data, deep learning-based methods have dominated in the nucleus segmentation task [23]. U-Net [24] and Mask R-CNN [25] are the two main deep learning frameworks used in nucleus segmentation. Both frameworks achieved top performances in Kaggle's 2018 Data Science Bowl [26]–[29]. U-Net is a U-shaped convolutional network widely used in bioimage and medical image segmentation tasks. This network uses skip connections to combine features at different resolutions to generate accurate pixel-level predictions, but it is not instance-aware. To adapt U-Net for instance segmentation, auxiliary tasks need to be added to learn features such as nuclei contours [30], overlapping borders [28], or within-nuclei locations [29] to separate touching nuclei instances. Then, some heavy postprocessing procedures are used to compose or fuse all pixel-level predictions to produce the segmentation of individual nuclei. Here, U-Net is used for pixel classification, and it has been shown that a comparable performance can be achieved when using a stacked random forest classifier as the classification backend [31]. A mask region-based convolutional neural network (Mask R-CNN), on the other hand, is an instance-aware framework and directly operates on a single-nucleus level. This tool predicts bounding boxes for all nuclei in an image and then segments the nuclei inside the predicted boxes. A Mask R-CNN can directly produce the segmentation of individual nuclei and, therefore, requires little or no postprocessing [27]. Moreover, a Mask R-CNN can detect individual nuclei more accurately than can U-Net [32], and it has a simpler and more flexible framework.

Despite performing well, both deep learning frameworks require training images to be fully annotated [33]. Images of nuclei can be acquired under a variety of conditions and have varied appearances depending on the cell type, tissue preparation method, staining method, imaging modality and imaging parameters. When there are no publicly available annotated datasets for a certain type of nucleus image, it is very time consuming to create fully annotated images manually, especially if the nuclei are densely packed and depict ambiguous overlapping. In contrast, it is relatively easy and quick to label representative exemplars (i.e., partial labeling) in the image. In creating a fully annotated image, it is less labor-intensive to refine or correct machine-generated annotations than to create annotations from scratch.

We extend the Mask R-CNN framework to robustly handle partially labeled training images. We adapt the self-attention mechanism [34], [35] to model long-range dependencies between nuclei. Inspired by the observed attention map (Fig. 1), we built an additional module to predict the centers of nuclei and bounding boxes around the centers, effectively propagating the partial annotations to the whole image. The predicted bounding boxes, after nonmaximum suppression, were merged with the other bounding box proposals generated by a Mask R-CNN and then sent to the standard bounding box-regression and binary mask-generation modules. We evaluated our method on a public dataset with 1/4 of its original annotations. The results indicate that our method can handle partially labeled training datasets stably compared to the baseline framework, achieving a level of performance comparable to that from training with full annotation. We show that our method is also useful for generating full annotation for new datasets. The source code is available at https://github.com/feng-lab/nuclei.

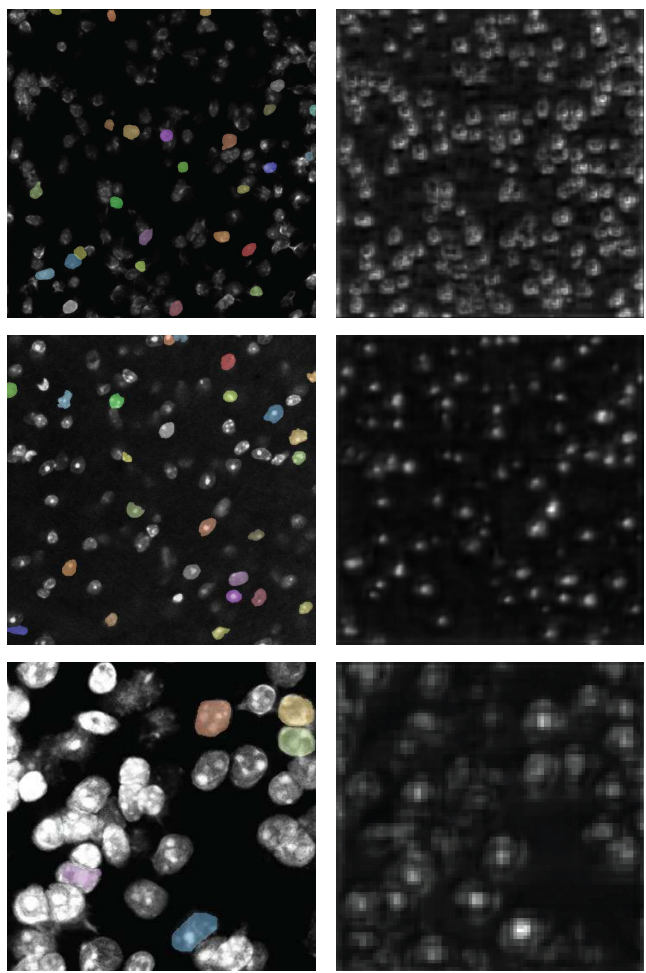

**FIGURE 1.** An illustration of the learned attention maps. In each row, the first image shows nuclei with a few color-coded annotations. The second image shows the merged attention map for the annotated (colored) regions. The self-attention module successfully models long-range correlation in images and generates meaningful attention focus, i.e., the centers and edges of nuclei.

## II. METHODS

In the Mask R-CNN pipeline, the region proposal network (RPN) generates a large number of bounding box candidates (also named anchors) that densely cover the entire image and then calculates an objectness score for each candidate, which indicates the probability of this candidate containing a nucleus. During training, all bounding box candidates are matched to the ground-truth nucleus bounding boxes. Candidates that have a large intersection over union (IoU), i.e., a high overlap ratio, with any ground-truth box are considered positive samples and have an objectness score of 1, whereas candidates that have low IoU values with all ground-truth boxes are considered negative samples and have an objectness score of 0. The RPN learns to calculate the objectness score by sampling positive and negative boxes and then computing the cross-entropy loss. The RPN selects and refines bounding box candidates with high objectness scores and then sends those selected to the region convolutional network (R-CNN). The R-CNN further evaluates the objectness of the selected candidates with more relevant features using a similar strategy to RPN to sample positive and negative boxes. If the training dataset is only partially labeled, the pipeline cannot sample negative boxes correctly because a candidate box might contain an unlabeled nucleus but have low IoU values with all labeled nuclei. These false negatives, if sampled by an RPN or R-CNN, could potentially impair the discrimination ability of the network and cause low detection recall. Therefore, we build a module to detect the bounding boxes of all potential nuclei in an image based on their similarity with the labeled nuclei. The detected boxes serve two purposes: 1) during training, they are used to prevent the RPN and R-CNN from sampling false negative boxes, and 2) during inference, they are used as "strong" proposals to improve the detection recall.

We used the self-attention mechanism [34] to capture the similarity between nuclei. The self-attention module calculates a response at a position as a weighted sum of the features at all positions, thereby efficiently modeling long-range correlations. The self-attention mechanism has improved the performance of convolutional networks on many vision tasks, including instance segmentation. We adopted the decomposed self-attention module [35] because it is more memory-efficient for handling large images. Specifically, for image feature x with $C$ channels and $N$ spatial locations ($x \in R^{C \times N}$), the decomposed self-attention enhanced feature is calculated as $x_{att} = x + W_f \left( \left( S \left( W_b x, C \right) \left( W_v x \right)^T \right)^T S \left( W_c x, N \right) \right)$, where $W_f \in R^{C \times C_v}$, $W_c \in R^{C_b \times C}$, $W_b \in R^{C_b \times C}$, and $W_v \in R^{C_v \times C}$ are parameter matrices implemented as $1 \times 1$ convolutions. Here, $S \left( W_b x, C \right)$ indicates channelwise softmax normalization, and $S \left( W_c x, N \right)$ indicates locationwise softmax normalization. $C_b$ and $C_v$ are channel counts of the basis and value, respectively. We use $C_b = C_v = 64$ in all our experiments. Fig. 1 shows the visualization of the attention map on the first-level feature of the feature pyramid network (fpn1) [36]. When we query the attention focus of the labeled nuclei, the self-attention module successfully captures the semantically related regions (i.e., all nuclei in the image) and generates a meaningful focus on the centers and edges of nuclei.

Inspired by the attention map, we build a simple partial annotation-handling module to predict nuclei centers and bounding boxes around centers. Fig. 2 illustrates the architecture of the Mask R-CNN with the proposed module. In brief, the Mask R-CNN is composed of a backbone network for generating multiscale features, an RPN module for generating bounding box proposals, an R-CNN head for evaluating and refining bounding box proposals, and a Mask head for generating binary masks, which, in this case, are the segmentations of all nuclei. Our proposed module uses the fpn1 feature because it has the highest spatial resolution among the features generated by the backbone network. It predicts the probability of being the center of a nucleus (i.e., the centerness score) at each spatial location of fpn1 and, for each nucleus center, a 4-dimensional bounding box regression result.

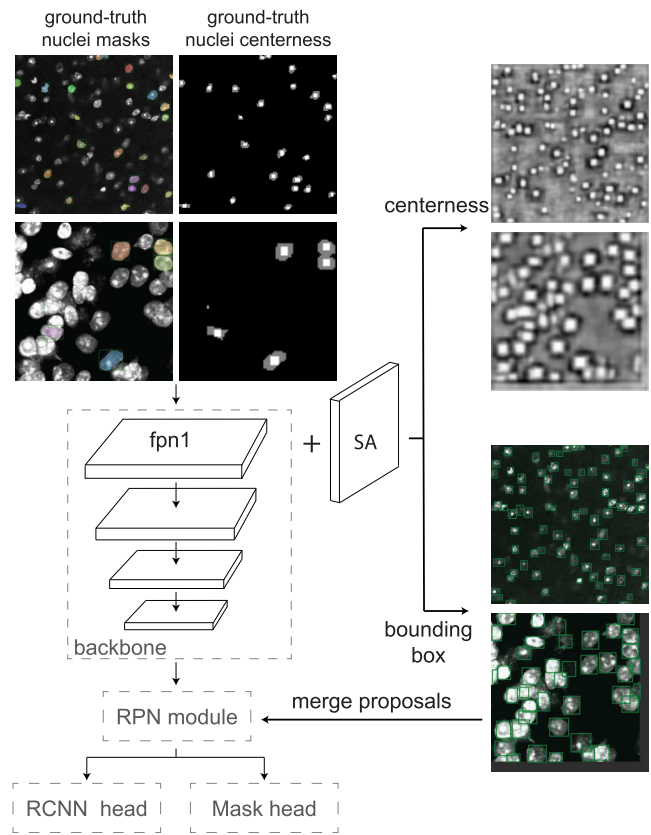

**FIGURE 2.** The proposed pipeline for learning from partially labeled exemplars. With only partially labeled ground-truth, we design a task to learn and predict a centerness score and a bounding box for each spatial location of fpn1, which is the first feature level of the backbone network. The predicted bounding boxes capture most of the nuclei in the image and are used to improve training and inference. Backbone, RPN module, RCNN head, and Mask head are standard Mask R-CNN components. SA is the decomposed self-attention module.

Given a labeled nucleus instance with binary mask M and bounding box $b = [x_c, y_c, w, h]$, $(x_c, y_c)$ is the center of the box and $w, h$ are the box width and height, respectively. We project both its binary mask and bounding box to fpn1 and obtain the projected (downsampled) binary mask $M^p$ and bounding box $b^p = \left[ x_c^p, y_c^p, w^p, h^p \right]$. We define the center of the nucleus bounding box as $b_c^p = \left[ x_c^p, y_c^p, 0.3 \times w^p, 0.3 \times h^p \right]$. In the ground-truth nucleus centerness map, we assign all pixels in $b_c^p$ to one (positive), assign all pixels in $M^p$ but not in $b_c^p$ to zero (negative), and assign all other pixels to -1 (ignored). Examples of ground-truth nucleus centerness maps are illustrated in Fig. 2 top. Positive, negative, and ignored pixels are shown in white, gray and black, respectively. For each positive pixel location $l^p$ with a ground-truth bounding box $b$, we empirically create an anchor bounding box $b_a = [x_l, y_l, \overline{w}, \overline{h}]$, where $(x_l, y_l)$ is the coordinate of $l^p$ in the image space and $\overline{w}, \overline{h}$ are the median box width and height in the training dataset, respectively.

During training, we randomly sample 128 pixels in the ground-truth nuclei center map to compute the cross-entropy loss $L_{centerness}$ of a batch, in which the sampled positive and negative pixels have a ratio of up to 1:1. For the sampled positive pixels, we use the smooth $L_1$ loss defined in [37] to calculate their bounding box regression loss $L_{centerbox}$. The overall loss used in training is

$$L = L_{centerness} + L_{centerbox} + L_{maskrcnn}, \quad (1)$$

where $L_{maskrcnn}$ is the loss used in the original Mask R-CNN pipeline.

The centerness score and bounding boxes predicted by our module can be visualized in the right bottom of Fig. 2. To obtain the predicted bounding boxes, we binarize the predicted centerness map with a threshold of 0.95. Then, for each connected component in the binarized centerness map, we choose the pixel with the highest centerness score and generate its bounding box using its anchor bounding box and the regression result. During training, the predicted bounding boxes are sent to the RPN and the R-CNN head to prevent them from sampling false negatives during loss calculation. During inference, the predicted bounding boxes are treated as "high score" candidates and are directly sent to the R-CNN for refinement.

## III. RESULTS

We evaluated the proposed method on the nucleus dataset published in [31]. The dataset contains fully annotated DAPI-stained cell nuclei, and we refer to it herein as DH100. To create a partially labeled dataset, we randomly removed 75% of the annotations from the training set and named the resulting dataset DH25. DH100 contains 60 images with 2120 nuclei annotations, while DH25 contains the same 60 images with 525 annotations. The baseline is a Mask R-CNN with a ResNeXt-101-64x4d backbone [38]. We trained the baseline model on DH100 and our proposed model on both DH100 and DH25 and then evaluated their performance on the fully annotated test set, which contains 23 images with 951 nuclei annotations. As the baseline model is not designed to handle partial annotation, it is not appropriate to directly train it on DH25. Instead we trained it on a dataset that contains a random crop (25%) of each image. This dataset, named DH25crop, is fully annotated and contains 676 nuclei annotations. Example training images with annotations from DH100, DH25, and DH25crop are shown in Fig. 3. All models were evaluated based on the fully annotated test set, which contains 23 images with 951 nuclei annotations. We initialized all models with the weights of the model pretrained on the MS-COCO dataset. All models were trained for 1000 iterations on 4 NVIDIA Tesla V100 graphics processing units (GPUs). The batch size was 8. We use a fixed learning rate of 0.01 with weight decay and gradient norm clipping. No image augmentation was used, except for vertical and horizonal flipping. For postprocessing, we simply dilated the detected nuclei with a rectangular structuring element of size $3 \times 3$, as suggested by [27]. Other hyperparameters followed the latest release of the Mask R-CNN benchmark [39].

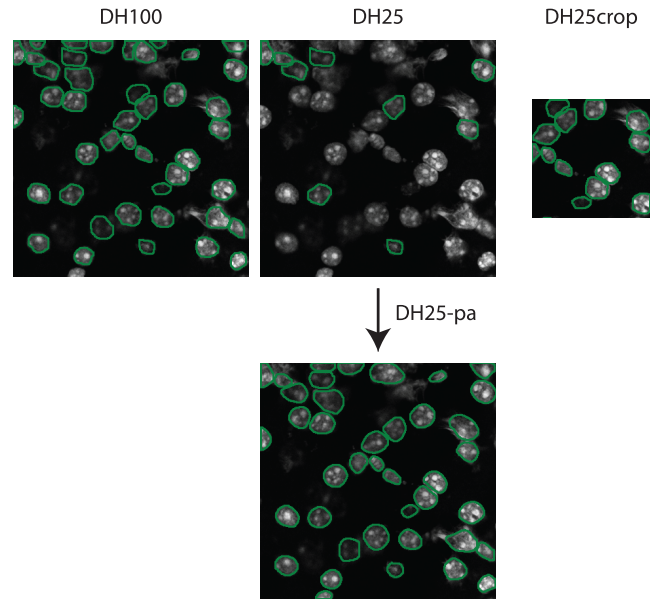

**FIGURE 3.** Examples of the training datasets used to evaluate our method. For each image in the original training set DH100, we randomly removed 75% of the annotations to get the corresponding training data in DH25 and randomly cropped 1/4 of the images to get the corresponding training data in DH25crop. The bottom row shows the nuclei predicted by our method.

### A. EVALUATION

The mean average precision (mAP) was used as the evaluation metric in the Kaggle competition [26]. mAP is the mean average precision at different thresholds of the IoU between the ground truth and predicted segmentation and is calculated as

$$\frac{1}{|\text{thresholds}|} \sum_{t=0.5:0.05:0.95} \frac{TP(t)}{TP(t) + FP(t) + FN(t)}, \quad (2)$$

where t is the IoU threshold and TP, FP, and FN are true positives, false positives, and false negatives, respectively.

The aggregated Jaccard index (AJI) [33], [40] is another commonly used metric for evaluating the segmentation performance. Formally, the AJI is defined as

$$AJI = \frac{\sum_{j=1}^{|G|} \left| r_j^G \cap r_*^S(j) \right|}{\sum_{j=1}^{|G|} \left| r_j^G \cup r_*^S(j) \right| + \sum_{l \in U} \left| r_l^S \right|}, \quad (3)$$

where $r^G$ represents the ground truth, $r^S$ is the predicted segmentation, $|G|$ is the number of ground-truth nuclei, $r_*^S(j)$ is the predicted nucleus that maximizes the IoU with the ground truth $r_j^G$ (with $r_*^S(i) \neq r_*^S(j)$ for $i \neq j$), and U is the set of predicted nuclei that are not matched to any ground truth.

Normalized coverage score (NCS) values were used in [31] to evaluate the matching quality between the ground truth and the predicted segmentation. The NCS is defined as

$$NCS = \frac{1}{\sum_{j=1}^{|G|} \left|r_j^G\right|} \sum_{j=1}^{|G|} \left|r_j^G\right| \max_{k=1,\ldots,|S|} IoU\left(r_j^G, r_k^S\right), \quad (4)$$

where $r^G$ is the ground truth, $r^S$ is the predicted segmentation, $|G|$ is the number of ground-truth nuclei, and $|S|$ is the number of predicted nuclei. The NCS is similar to the AJI, but it does not punish false-positive detections.

We use the mAP and AJI as our main metrics, and we also report the NCS, precision at an IoU of 0.5, and recall at an IoU of 0.5 for reference or comparison.

### B. OVERALL PERFORMANCE

We refer to the baseline Mask R-CNN model as "DH100-baseline" or "DH25crop-baseline", depending on the training dataset. Similarly, we refer to our model, which is the baseline Mask R-CNN plus our partial annotation-handling module, as "DH100-pa" or "DH25-pa". To examine the performance dynamics of these models as training progressed, we evaluated their performance at every 50100 iterations, which corresponded to approximately 613 epochs for this dataset. Fig. 4 shows the performances of all models as measured by the aforementioned metrics. We summarize the best mAP for each model, along with the AJI, in Table 1.

When the baseline model is directly trained on DH25, its detection recall is extremely low, while its detection precision, which is measured as $TP/(TP+FP)$, is not greatly affected (data not shown). This is consistent with our previous analysis indicating that sampling false negatives during training will cause low recall. Indeed, the baseline model works much better when trained on the cropped but fully annotated dataset DH25crop. But as shown in Fig. 4, there is still a large performance gap between "DH25crop-baseline" and "DH100-baseline". Compared to the baseline model trained on the full dataset ("DH100-baseline"), "DH25-baseline" shows consistently decreased mAP and AJI values, indicating the importance of having a large training dataset for the baseline model.

In contrast, our model "DH25-pa" shows comparable performance to that of "DH100-baseline", despite being trained with fewer annotations. This performance demonstrates that our proposed module can alleviate the aforementioned false-negative sampling problem and thus help the network learn from the partially annotated dataset. Additionally, note that the recall rate (as well as the NCS) of "DH25-pa" increases as the training progresses, indicating that our proposed module gradually learns to detect increasing numbers of nuclei in the unlabeled regions. Our "DH25-pa" (trained with 525 annotations) consistently outperforms "DH25crop-baseline" (trained with 676 annotations) in terms of both the mAP and AJI, showing that our partial annotation-handling

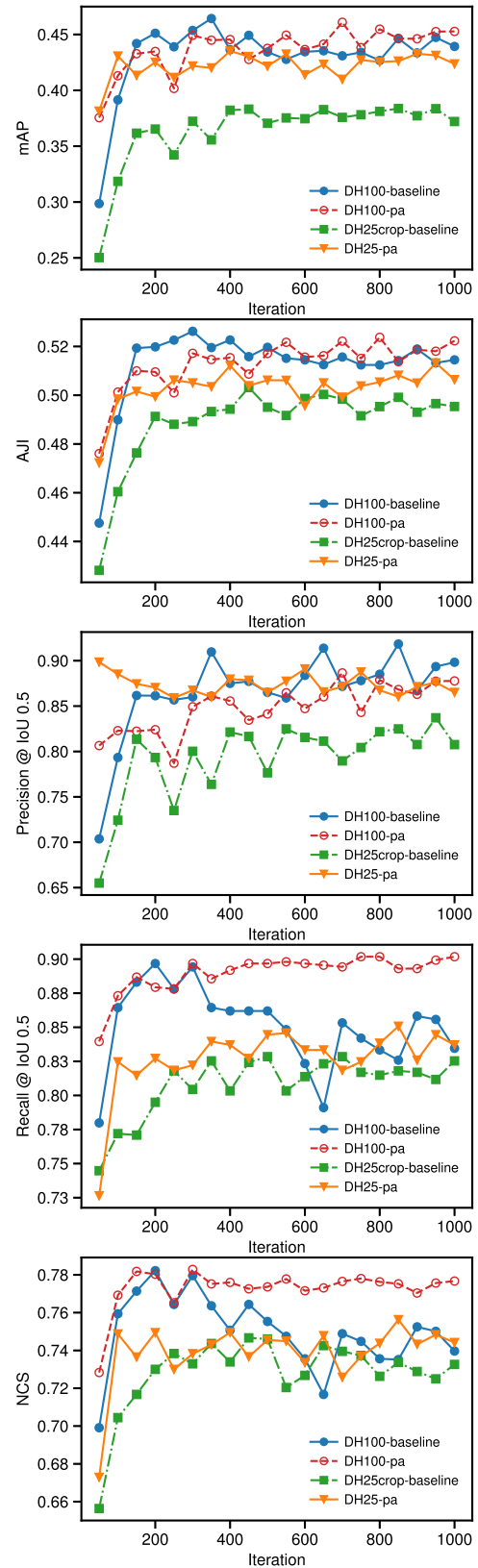

**FIGURE 4.** The performance of all models as measured by the mAP, AJI, precision at IoU of 0.5, recall at IoU of 0.5 and NCS.

module is effective in learning from a limited number of training examples. One visual result is illustrated in the bottom

**TABLE 1.** Best evaluation performance of each model.

| Model | Annotation (Type, No.) | mAP | AJI |
|---|---|---|---|
| DH100-baseline | full, 2120 | **0.464** | 0.520 |
| DH100-pa | full, 2120 | 0.461 | **0.522** |
| DH25crop-baseline | full, 676 | 0.384 | 0.499 |
| DH25crop-pa | full, 676 | 0.393 | 0.504 |
| DH25-pa | partial, 525 | **0.435** | **0.512** |

The best evaluation scores on each dataset are indicated by bold text.

**TABLE 2.** Size and speed of each model.

| Model | No. parameters (millions) | Train time (s/iter) | Inference time (s/im) | Train memory (MB) |
|---|---|---|---|---|
| DH100-baseline | 102.1 | 0.511 | 0.049 | 3288 |
| DH100-pa | 103.1 | 0.582 | 0.083 | 3363 |
| DH25crop-baseline | 102.1 | 0.326 | 0.045 | 2840 |
| DH25crop-pa | 103.1 | 0.370 | 0.054 | 2859 |
| DH25-pa | 103.1 | 0.464 | 0.084 | 3163 |

of Fig. 3. Furthermore, all of our models, even those trained with fewer annotations, show better performances in terms of the NCS than do those (approximately 0.6) reported in [31].

Although our model is designed to handle partially annotated datasets, it does not degrade the performance of the baseline framework when used with a fully annotated dataset. As shown in Fig. 4, the "DH100-pa" model has a close performance to that of "DH100-baseline". The former also shows a slightly superior performance in terms of the NCS and recall, indicating that the module we added can generate high-quality nucleus proposals, which could potentially be used to further improve the detection framework. Also, it is worth noting that as the training progresses, the performance of "DH100-pa" becomes more stable and shows fewer signs of over-fitting than does "DH100-baseline".

To further demonstrate the advantage (or necessity) of having sparse annotations (DH25) over having full annotations on a smaller area (DH25crop), we trained our method on the DH25crop ("DH25crop-pa") and compared it to the baseline model. Similar to the DH100 case, "DH25crop-pa" shows limited improvement of the overall performance over that of "DH25crop-baseline" (Table 1). This result indicates that our additional learning task itself is not sufficient for achieving good performance. It is equally important to have unlabeled regions to help in the training. Our method can effectively exploit the unlabeled regions through the attention mechanism and thereby improve the detection performance. In this sense, our method can be considered a semisupervised learning method that utilizes both labeled and unlabeled data.

We report the size and speed of each model in Table 2. Compared to the baseline model "DH100-baseline", our proposed model "DH25-pa" has a slightly larger number of parameters (<1% increase). While the inference time of our model is increased because of our additional module, the training time and the GPU memory consumption are decreased compared to the baseline model, possibly due to the reduced number of training annotations.

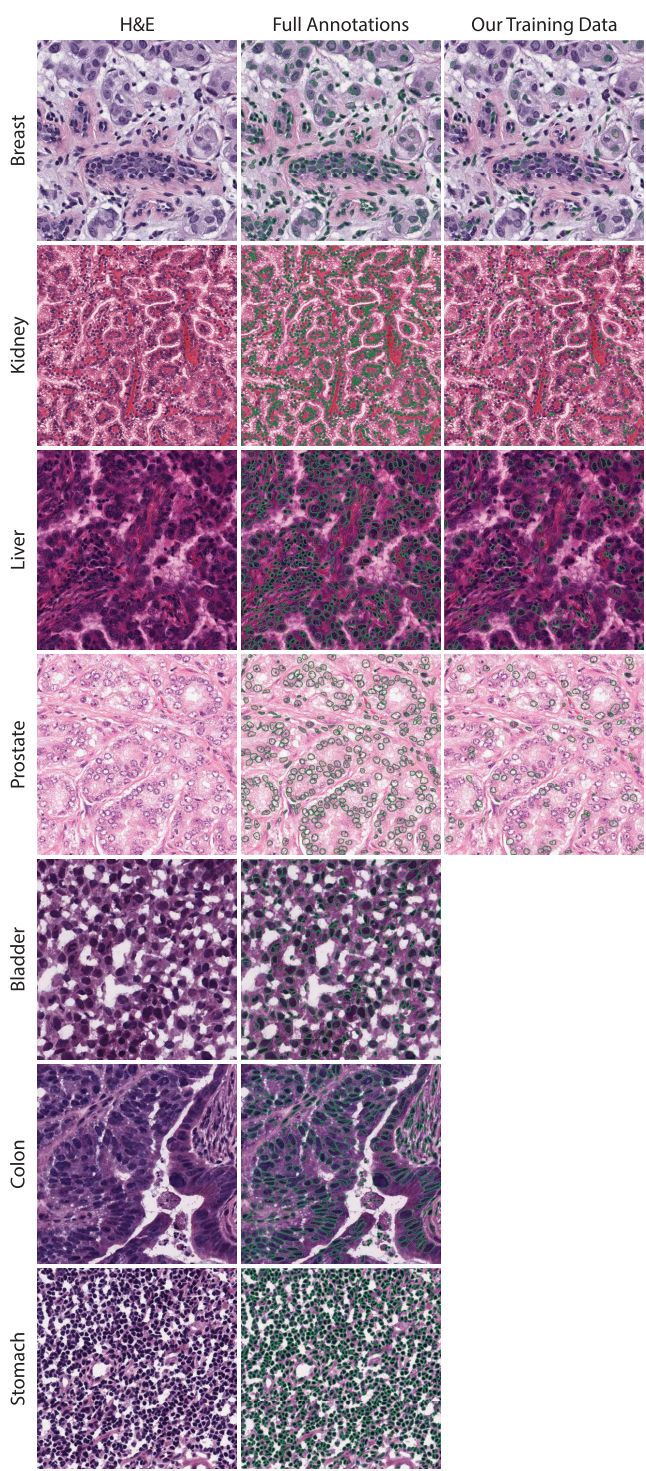

**FIGURE 5.** Examples of histopathology images for different organs (rows) showing various nuclear appearances. What makes this task even more challenging is that test images from the bladder, colon, and stomach are not represented in the training set. Our method is trained with only 25% of the full annotation.

### C. HISTOPATHOLOGY IMAGES
To demonstrate the generality of our method, we evaluated it on hematoxylin and eosin (H&E)-stained histopathology datasets presented in [33] and [40]. The first dataset consists of 30 annotated histology images of patients from several hospitals [40]. These images come from 7 different organs and represent different cancer types. An illustration of the variability of tissue appearances and their nuclei observed in this dataset is shown in Fig. 5. We use exactly the same subset of images (14 images) for testing as used in [40] and [33] to benchmark against their methods. After randomly removing 75% of the annotations from the 16 training images, we trained our network for 2000 iterations with a batch size of 4. Table 3 shows the overall performances of the different methods as well as their performances on individual test images. At the individual image level, there were no statistically significant performance differences between our method and the other two methods that use fully annotated training data (Wilcoxon signed-rank test). Notably, our method achieved this performance not only with fewer nuclei annotations (25%) but also with fewer data augmentation techniques. During training, our method uses only vertical and horizontal flipping, while the methods presented in [40] and [33] employ complex data augmentation techniques such as rotation, blurring, image deforming, and color deconvolution.

**TABLE 3.** Performance comparison of different methods on individual histopathology test images.

| Organ | Image | AJI ours | DIST [33] | CNN3 [40] |
|---|---|---|---|---|
| Breast | 1 | 0.5027 | 0.5334 | 0.4974 |
| | 2 | 0.5784 | 0.5884 | 0.5796 |
| Kidney | 1 | 0.5271 | 0.5648 | 0.4792 |
| | 2 | 0.4727 | 0.5420 | 0.6672 |
| Liver | 1 | 0.5150 | 0.5466 | 0.5175 |
| | 2 | 0.4820 | 0.4432 | 0.5148 |
| Prostate | 1 | 0.4640 | 0.6273 | 0.4914 |
| | 2 | 0.5708 | 0.6294 | 0.3761 |
| Bladder | 1 | 0.6313 | 0.6475 | 0.5465 |
| | 2 | 0.4127 | 0.5467 | 0.4968 |
| Colon | 1 | 0.4359 | 0.4240 | 0.4891 |
| | 2 | 0.5008 | 0.4484 | 0.5692 |
| Stomach | 1 | 0.6367 | 0.6408 | 0.4538 |
| | 2 | 0.6453 | 0.6550 | 0.4378 |
| Overall | | 0.5353 | 0.5598 | 0.5083 |

**TABLE 4.** Performance of different methods on the breast cancer slides.

| Method | Annotation (Type, No.) | AJI |
|---|---|---|
| CNN3 [40] | full, 13372 | 0.5385 |
| DIST [33] | full, 5577 | 0.5853 |
| ours | partial, 1425 | 0.5497 |

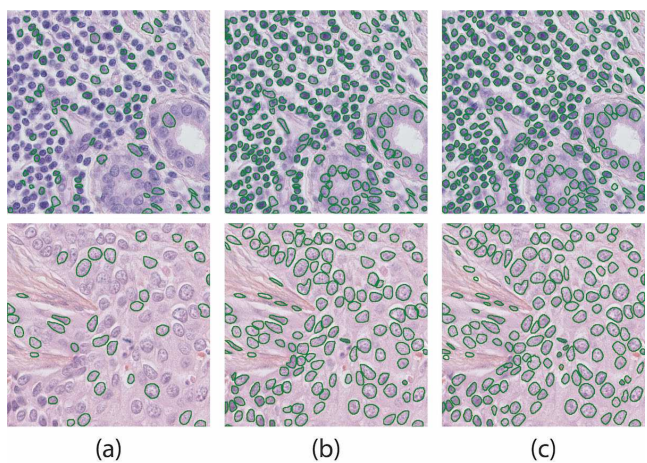

**FIGURE 6.** Results from the histopathology images show that our method successfully propagates partial annotations to the whole image. (a) Annotations used for training. (b) Ground truth. (c) Nuclei predicted by our method.

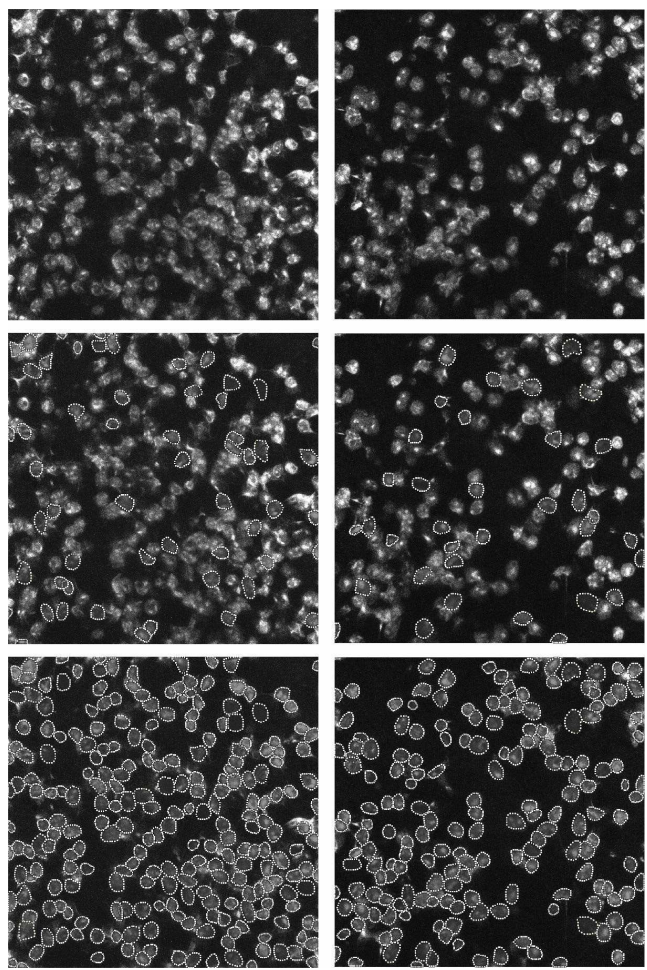

**FIGURE 7.** Our method used as a computer-assisted annotating tool. The first row shows the example images to be annotated. The second row shows the partial annotations created by a human annotator. It is difficult for human annotators to annotate all nuclei in the images because of the high density and weak nuclei edges. The third row shows the annotations generated by our method after training. The generated annotations are presented as natural cubic splines with control points for further manual editing.

The second dataset consists of 50 H&E-stained images from 11 patients with triple negative breast cancer (TNBC). The top performing model for breast cancer samples in [33] used 54 images (4 of which come from [40]) for training and 2 images from [40] for testing. The training images contained 5577 fully annotated nuclei, and the testing images contained 710 nuclei. We used the same training images here, except that we randomly removed 75% of the annotations from the training set so that our method was trained with only 1425 partially annotated nuclei. We trained our network for 1000 iterations with a batch size of 8. To compare the performance of our method with those of the methods presented in [33] and [40], we performed inference on the same test set and evaluated the accuracy using the AJI. The results are shown in Table 4. Compared to the methods that use full annotation, our method reduces the requirements for annotation and still achieves a comparable performance. Importantly, our method can outperform the method in [40] with a much smaller training set. We also perform inferences on the training images and compare the results with the ground-truth annotations (Fig. 6). Visual inspection shows that our method successfully learns from the annotated exemplars and propagates the partial annotations to the whole image.

### D. APPLICATION

In this section, we show that our method can also be used as a computer-assisted annotating tool to create fully annotated datasets. In brief, 20-$\mu$m-thick coronal sections of mouse brains were counterstained with DAPI and imaged with an LSM 780 confocal microscope (Carl Zeiss Microscopy). As shown in Fig. 7, it is difficult to annotate all nuclei in these images because of the high nucleus density and ambiguous nuclei overlapping, but it is relatively easy and fast to perform partial annotation. We trained our network on 26 partially annotated images (second row of Fig. 7) for 200 iterations. Then, we performed inferences on the same 26 images. As shown in the third row of Fig. 7, our method can robustly learn from the labeled nucleus examples and propagate high-quality annotations to the whole image. We converted the segmentation result into natural cubic splines with control points for further manual refinement.

## IV. CONCLUSION

The automatic and reliable characterization of cells in images is key to many biological applications. Recently, deep learning-based segmentation frameworks have greatly improved segmentation performances and have achieved state-of-the-art results in many competitions. However, most of these frameworks require a fully annotated training dataset. It is often difficult to find compatible, public datasets with annotations because of the large variability in microscopy images [33]. In this work, we show that deep convolutional networks can capture the semantic similarity between nuclei in the same image by visualizing the self-attention map. Therefore, it is possible to extend the current frameworks to handle partially annotated training data.

We propose a simple module to learn from a few exemplar annotations and then detect the potential unlabeled nuclei. We integrate this module into the Mask R-CNN framework and show that our proposed method can robustly learn from a partially labeled dataset. Our method uses fewer training data than does its predecessor and achieves a comparable performance to that of the baseline method; therefore, it could greatly reduce the time needed for creating training datasets.

Our method utilizes the similarity between the labeled regions and the unlabeled regions, it is worth to study the minimum required annotations for different datasets in future work. The detection performance is slightly decreased when training with partial annotations because our network has fewer examples from which to learn. In future work, we will study this in detail to bridge the gap and compare our model with more state-of-the-art methods. One potential solution is to use data augmentation, which is not used in this work because it would complicate the analyses. However, it is difficult to find a good data augmentation strategy that will work well on different datasets [41]. Our method provides a better solution—when extreme segmentation accuracy is needed, our method can be used as a computer-assisted annotation tool to help create high-quality fully annotated training datasets.

It has been shown that when nuclei are correctly isolated, U-Net-based frameworks can generate better binary masks, probably because of the high-resolution features used in U-Net [32]. In this work, we performed nucleus prediction and regression on fpn1, whose spatial resolution (1/4) was lower than the image resolution. Problems may ensue if some of the nuclei to be detected are extremely small. In the future, our method could be extended to use some high-resolution networks [42] to further improve the detection and segmentation accuracy for nuclei with different sizes.

### ACKNOWLEDGMENT
The authors would like to thank M. Cicconet, N. Kumar, and T. Walter for providing the training datasets.

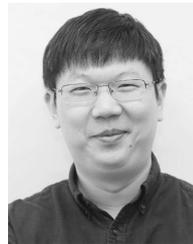

**LINQING FENG** received the B.S. and Ph.D. degrees in biomedical engineering from Zhejiang University, China, in 2006 and 2011, respectively. From 2011 to 2016, he was a Postdoctoral Researcher with the Dr. J. Kim's Laboratory, Korea Institute of Science and Technology (KIST), South Korea. Since 2016, he has been a Senior Researcher with the Center for Functional Connectomics, KIST. His current research interests include bioimage informatics, image-based brain connectivity mapping, neuroinformatics, and big data visualization. He was a recipient of The International Society for Computational Biology (ISCB) Travel Award, in 2012.

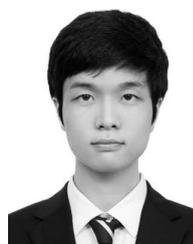

**JUN HO SONG** received the B.Eng. and M.Res. degrees in biomedical engineering from the Imperial College of Science, Technology, and Medicine, London, U.K., in 2014 and 2015, respectively. Since 2016, he has been a Research Scientist with KAIST, Daejeon, South Korea, and KIST, Seoul, South Korea. His current research interests include sensory/behavioral neuroscience and bioimage processing.

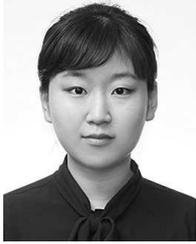

**JIWON KIM** was born in 1994. She received the B.S. degree in life sciences from Korea University, Seoul, South Korea, in 2017, and the M.S. degree in bio-medical science and technology (neuroscience) from the Korea Institute of Science and Technology (KIST) School, University of Science and Technology, Seoul, in 2019. She is currently an Intern with the Dr. J. Kim's Laboratory, KIST, Seoul. Her current research interests include the development of genetic tools for brain circuit mapping and the molecular profiling of neural cells in the mouse or human brain. She was a recipient of the 9th Federation of the Asian and Oceanian Physiological Societies Congress Young Scientist Travel Award, in 2019.

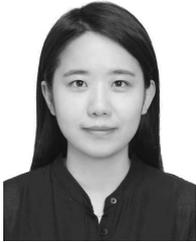

**SOOMIN JEONG** received the B.S. degree in molecular and cellular biology from the University of Illinois at Urbana–Champaign, IL, USA, in 2013, and the M.S. degree in medicine from Seoul National University, Seoul, South Korea, in 2018. She is currently pursuing the Ph.D. degree with the University of Wisconsin–Madison, WI, USA. In 2019, she was with the Korea Institute of Science and Technology (KIST) to propose an improved quantitative analysis of microscopical images. Her current research interests include understanding the molecular mechanisms and functional recovery in injured brain.

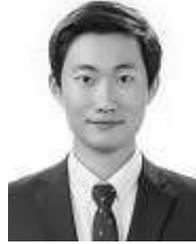

**JIN SUNG PARK** was born in 1992. He received the B.S. degree from the Department of Biology, Kyung Hee University, Seoul, South Korea, in 2017, and the M.S. degree in bio-medical science and technology (neuroscience) from the Korea Institute of Science and Technology (KIST) School, University of Science and Technology, Seoul, in 2019. He is currently an Intern with the Dr. J. Kim's Laboratory, KIST, Seoul. His current research interest includes the development of genetic tools for brain circuit mapping.

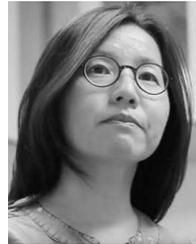

**JINHYUN KIM** received the B.S. and M.S. degrees in biology from Sung Kyun Kwan University, South Korea, in 1995 and 1997, respectively, and the Ph.D. degree in neuroscience from the Max-Planck-Institute for Medical Research, Heidelberg, Germany, in 2001. After her Ph.D., she did her 5-year postdoctoral work with the National Institutes of Health, USA, from 2002 to 2007, and a Research Specialist with the Howard Hughes Medical Institute, Janelia Research Campus, USA, from 2008 to 2010. Since 2011, she has been appointed by the Korea Institute of Science and Technology participating in a World-Class-Institute launched by Korean Government. In 2015, she became the Director of the Center for Functional Connectomics.

● ● ●